\documentclass[aip,apl,preprint]{revtex4-1}

\usepackage{graphicx}
\usepackage{dcolumn}
\usepackage{bm}
\usepackage{amssymb}
\usepackage{wrapfig}

\begin{document}

\title{A micro-sized parametric spin-wave amplifier}

\author{T. Br\"acher}
\affiliation{Fachbereich Physik and Forschungszentrum OPTIMAS, Technische Universit\"at
Kaiserslautern, D-67663 Kaiserslautern, Germany}
\author{F. Heussner}
\affiliation{Fachbereich Physik and Forschungszentrum OPTIMAS, Technische Universit\"at
Kaiserslautern, D-67663 Kaiserslautern, Germany}
\author{P. Pirro}
\affiliation{Fachbereich Physik and Forschungszentrum OPTIMAS, Technische Universit\"at
Kaiserslautern, D-67663 Kaiserslautern, Germany}
\author{T. Fischer}
\affiliation{Fachbereich Physik and Forschungszentrum OPTIMAS, Technische Universit\"at
Kaiserslautern, D-67663 Kaiserslautern, Germany}
\author{M. Geilen}
\affiliation{Fachbereich Physik and Forschungszentrum OPTIMAS, Technische Universit\"at
Kaiserslautern, D-67663 Kaiserslautern, Germany}
\author{B. Heinz}
\affiliation{Fachbereich Physik and Forschungszentrum OPTIMAS, Technische Universit\"at
Kaiserslautern, D-67663 Kaiserslautern, Germany}
\author{B. L\"agel}
\affiliation{Fachbereich Physik and Forschungszentrum OPTIMAS, Technische Universit\"at
Kaiserslautern, D-67663 Kaiserslautern, Germany}
\author{A. A. Serga}
\affiliation{Fachbereich Physik and Forschungszentrum OPTIMAS, Technische Universit\"at
Kaiserslautern, D-67663 Kaiserslautern, Germany}
\author{B. Hillebrands}
\affiliation{Fachbereich Physik and Forschungszentrum OPTIMAS, Technische Universit\"at
Kaiserslautern, D-67663 Kaiserslautern, Germany}

\date{\today}

\begin{abstract}
We present the experimental observation of the localized amplification of externally excited spin waves in a transversely in-plane magnetized Ni$_{81}$Fe$_{19}$ magnonic waveguide by means of parallel pumping. By employing microfocussed Brillouin light scattering spectroscopy, we analyze the dependency of the amplification on the applied pumping power and on the delay between the input spin-wave packet and the pumping pulse. We show that there are two different operation regimes: At large pumping powers, the spin-wave packet needs to enter the amplifier before the pumping is switched on in order to be amplified while at low powers the spin-wave packet can arrive at any time during the pumping pulse.
\end{abstract}

\pacs{}

\maketitle

The field of magnonics\cite{Lenk-2011-1, Chumak-2010-1} deals with the processing of information through the utilization of spin waves and their quanta, magnons. The prospect to transmit spin currents through electrical insulators in the form of spin waves\cite{Stamps-2014-1} and the creation of a wave-based logic\cite{Schneider-2008-1, Chumak-2014-1} where the information is encoded into their amplitude and phase\cite{Pirro-2011-1, Lee-2008-1} have led to an increased research effort in this field. Although first experiments using materials which feature a small spin-wave damping have already shown promising results,\cite{Sebastian-2012-1, Pirro-2014-1} one of the main challenges for the field of magnonics is still the limited lifetime of spin waves. Parallel parametric amplification, or in short, parallel pumping, is one possible way to overcome this issue.\cite{Chumak-2010-1} Parallel pumping describes the splitting of microwave photons into pairs of magnons at half the photon frequency.\cite{Schloemann-1960-1} Recently, it was demonstrated that this technique can be used to amplify externally excited spin waves in a microstructured Ni$_{81}$Fe$_{19}$ magnonic waveguide.\cite{Braecher-2014-2} This allows for a significant increase in their effective lifetime and, thus, propagation distance. Still, parallel pumping also amplifies thermal spin waves, leading to parametric generation. If the pumping is applied along the whole waveguide, the coexistence of the parametric amplification of the externally excited spin waves with the parametric generation reduces the efficiency of the amplification and makes an experimental separation of these two processes challenging.

\begin{figure}[b!]
	  \begin{center}
    \scalebox{1}{\includegraphics[width=7.3 cm, clip]{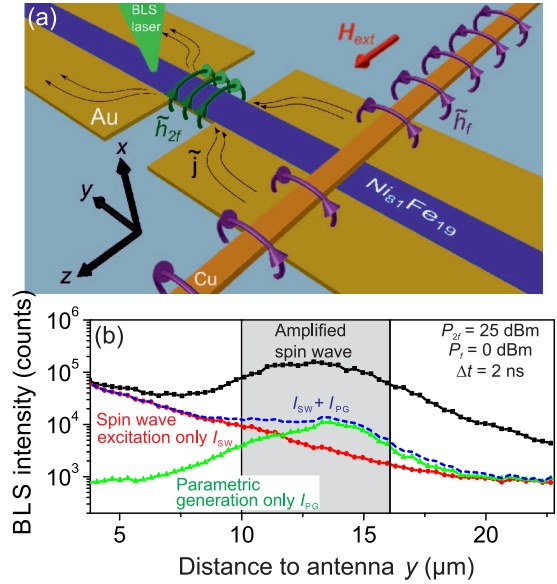}}
    \end{center}
	  \caption{\label{Figure1}(a) Sample schematic: A Ni$_{81}$Fe$_{19}$ waveguide lies on top of a microstrip transmission line. The dynamic Oersted field $\tilde{h}_f$ created by a microstrip antenna across the waveguide provides an external spin-wave excitation in the transversely in-plane magnetized waveguide. At a distance of $10\,\mu\mathrm{m}$ from this antenna, the transmission line features a narrowed region where the microwave current density $\tilde{j}$ and, consequently, the dynamic Oersted field $\tilde{h}_{2f}$ are enhanced, acting as the amplifier. (b) Logarithmic BLS intensity vs. distance from the antenna (integrated over the waveguide width) for different configurations. Red circles: only spin-wave excitation by the antenna ($I_\mathrm{SW}$). Green triangles: only parametric generation ($I_\mathrm{PG}$). Black squares: spin-wave amplification. Dashed line: $I_\mathrm{SW}$ + $I_\mathrm{PG}$ for comparison. The shaded area marks the geometric extents of the amplifier.}
\end{figure}

Here, we report on the localized parallel parametric amplification of externally excited spin waves in a transversely in-plane magnetized Ni$_{81}$Fe$_{19}$ magnonic waveguide. In particular, we demonstrate how the parametrically generated magnons limit the efficiency of the amplification. We show that there are two possible regimes to use the parametric amplifier: In the high-power regime, the externally excited spin waves are only amplified if the pumping is switched on after the spin waves have entered the amplifier region while, in the low-power regime, the spin-wave packet can enter at an arbitrary time within the pumping pulse and its amplification is always observed.

The investigated sample is sketched in Fig.~\ref{Figure1}~(a) and consists of a Ni$_{81}$Fe$_{19}$ magnonic waveguide (length $L_\mathrm{s} = 75\,\mu\mathrm{m}$, width $w_\mathrm{s} = 4\,\mu\mathrm{m}$, thickness $d_\mathrm{s} = 40\,\mathrm{nm}$) which has been patterned on top of a $d = 250\,\mathrm{nm}$ thick Au microstrip transmission line. The Au and the Py are separated by a $d_\mathrm{HSQ} = 250\,\mathrm{nm}$ thick insulation layer made from Hydrosilesquioxane. In an additional step, a Cu microstrip antenna ($w_\mathrm{a} = 2\,\mu\mathrm{m}$, $d_\mathrm{a} = 500\,\mathrm{nm}$) has been patterned across the magnonic waveguide. The waveguide is magnetized transversely in-plane by an applied magnetic bias field of $\mu_0 H_\mathrm{ext} = 47\,\mathrm{mT}$. As a result, the dynamic Oersted field $\tilde{h}_f$ created by the microstrip antenna is perpendicular to $H_\mathrm{ext}$ and, thus, acts as an effective spin-wave excitation source.\cite{Demidov-2009-1} The regular width of the transmission line is $w_1 = 20\,\mu\mathrm{m}$. At a distance of $l_1 = 10\,\mu\mathrm{m}$ from the microstrip antenna, the transmission line features a $l_2 = 6\,\mu\mathrm{m}$ long narrowed region with a width of $w_2 = 4\,\mu\mathrm{m}$. In this narrowed region, the current density is larger than in the wider parts of the transmission line and, thus, the local dynamic Oersted field is enhanced. As it has been demonstrated in Ref. \onlinecite{Braecher-2014-1}, the combination of this local enhancement with the strong dependency of the parametric amplification on the microwave field can be used to create a local parallel parametric amplifier\cite{Schloemann-1960-1} for spin waves in a transversely in-plane magnetized magnonic waveguide. In the following, this region is, therefore, addressed as the \textit{amplifier}. 

As mentioned above, parallel pumping describes the creation of magnon pairs at frequency $f$ out of microwave photons with frequency $2f$ and relies on the coupling of the microwave field to the longitudinal component of the dynamic magnetization. The latter arises from the ellipticity of precession.\cite{Melkov-1996-1} The magnon creation opposes the magnon damping and, if the number of magnons injected per unit time exceeds the losses, the parametric instability threshold is reached and the number of magnons exponentially grows over time.\cite{Lvov-1994-1} In general, parallel pumping can interact with externally excited, coherent magnons and can be used to amplify these spin waves.\cite{Braecher-2014-2, Smith-2007-1} On the other hand, it can also interact with thermal magnons present in the magnonic waveguide and, if the instability threshold of these spin waves is reached, one refers to parametric generation.\cite{Braecher-2011-1} 

Figure \ref{Figure1}~(b) shows the functionality of the localized spin-wave amplifier: Spin waves are excited at the antenna by a $\tau_f = 15\,\mathrm{ns}$ long microwave pulse with a carrier frequency of $f=6\,\mathrm{GHz}$ and an excitation power of $P_f =0\,\mathrm{dBm}$ and propagate towards the amplifier. Just after the spin-wave packet has entered the amplifier, a $\tau_{2f} = 25\,\mathrm{ns}$ long microwave pulse with a carrier frequency of $2f = 12\,\mathrm{GHz}$ and a power of $P_{2f} = 25\,\mathrm{dBm}$ is applied to the pumping transmission line. Since the microwaves are applied by two different, not phase-locked microwave generators, the phase between the excitation and the amplification is random and not fixed in time. This corresponds to the real operation conditions of the majority of amplifiers, where the phase of the input waves is unknown. The applied pumping power $P_{2f}$ is larger than the power needed to overcome the parametric instability threshold in the center of the amplifier, which has been determined to be $P_{2f,\mathrm{th}}=20\,\mathrm{dBm}$ in an additional measurement with longer microwave pulses (not shown), by $5\,\mathrm{dB}$. Utilizing microfocussed Brillouin Light Scattering spectroscopy (BLS)\cite{Demidov-2004-1} the total spin-wave intensity (integrated across the width of the waveguide) at $f = 6\,\mathrm{GHz}$ is studied as a function of the distance to the microstrip antenna. The scenario described above (i.e. spin waves being emitted and interacting with the pumping field) is shown by the black curve in Fig.~\ref{Figure1}~(b). As can be seen from the figure, the spin waves decay initially and are amplified once they enter the effective amplifier region. The geometric confinement of the amplifier is indicated by the shaded area in Fig.~\ref{Figure1}~(b). However, the effective size of the amplifier region, as has been demonstrated in Ref. \onlinecite{Braecher-2014-1}, is larger due to the redistribution of the current in the wider part of the transmission line. After the spin waves have passed the center of the amplifier, where the amplification is strongest, the spin-wave intensity decreases. Once the spin waves leave the effective amplifier region they decay without amplification. For comparison, the red curve shows the BLS intensity $I_\mathrm{SW}$ if spin waves are excited at the antenna and no pumping is present during the spin-wave propagation. The green curve shows the BLS intensity $I_\mathrm{PG}$ arising from parametric generation only, i.e., if the pumping pulse is applied and no spin waves are excited at the antenna. For reference, the blue curve shows the sum of the two individual signals $I_\mathrm{SW}+I_\mathrm{PG}$ which would arise if the spin-wave packet and the pumping would not interact with each other. Comparing the black and the blue curve, it is obvious that a strong amplification takes place and, as a consequence, the spin waves travel significantly farther than without amplification while the small number of parametrically generated spin waves is negligible at distances larger than $20\,\mu\mathrm{m}$ from the antenna. The latter is a general benefit of the localized amplification in comparison to the non-local amplification studied in Ref. \onlinecite{Braecher-2014-2}.

\begin{figure}[t!]
	  \begin{center}
    \scalebox{1}{\includegraphics[width=7.5 cm, clip]{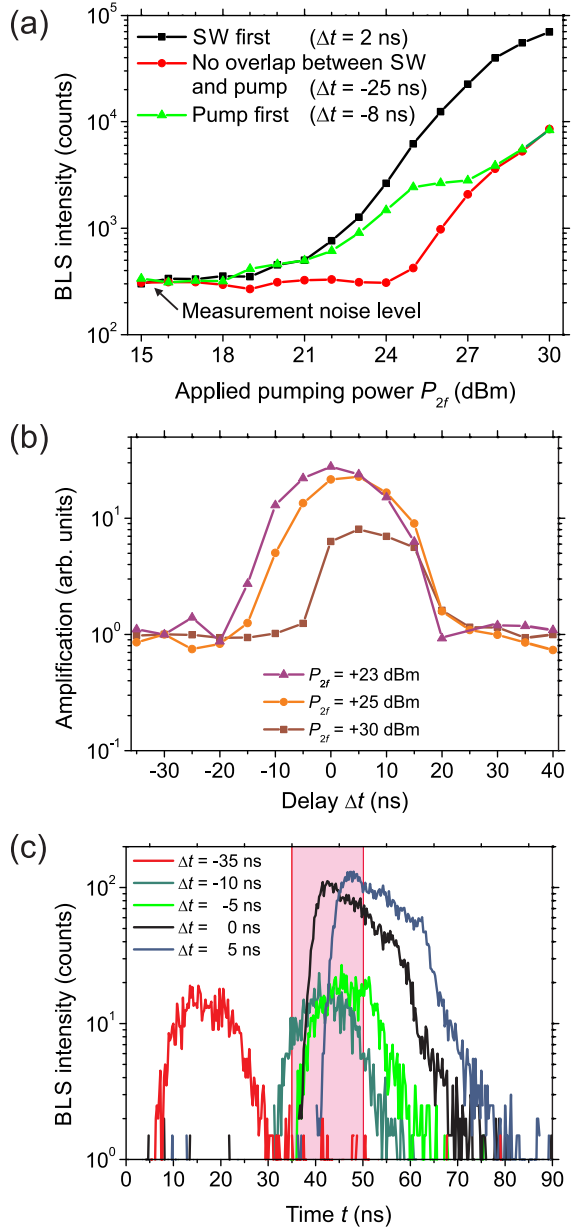}}
    \end{center}
	  \caption{\label{Figure2}(a) Influence of the applied pumping power on the measured BLS intensity $9\,\mu\mathrm{m}$ behind the amplifier ($25\,\mu\mathrm{m}$ from the microstrip antenna) for different pulse configurations. Black squares: Spin wave arrives first ($\Delta t > 0$). Red dots: Spin-wave packet and pumping pulse have no temporal overlap. Green triangles: Pumping pulse arrives first ($\Delta t < 0$). (b) Influence of delay $\Delta t$ for different $P_{2f}$. (c) Time-resolved BLS intensity for different delays $\Delta t$ for $P_{2f} = 30\,\mathrm{dBm}$. The shaded area indicates the extents of the externally excited spin-wave packet in time.}
\end{figure}

To study the efficiency of the amplifier, the amount of amplified as well as generated spin waves at a distance of about $25\,\mu\mathrm{m}$ from the antenna is analyzed as a function of the applied pumping power $P_{2f}$ and the delay $\Delta t = t_{0,2f} - t_{0,f}$, where $t_0$ denotes the starting time of the respective pulses. The $P_{2f}$ dependency for three different delays is shown in Fig.~\ref{Figure2}~(a): Black: The spin-wave packet enters the amplifier shortly before the pumping is applied. Red: The spin-wave and the pumping pulse are completely displaced in time, i.e. there is no temporal overlap and, thus, no amplification. Green: The spin-wave packet enters the amplifier after the pumping pulse is applied. 

From Fig. \ref{Figure1}~(b) it can be extrapolated that the spin waves have decayed to about $10\%$ of the noise level at the measurement point of Fig.~\ref{Figure2}~(a). Thus, the remaining BLS intensity at $P_{2f}<25\,\mathrm{dBm}$ in this scenario is mainly caused by the measurement noise. As mentioned above, the power needed to overcome the instability threshold has been determined to $P_{2f,\mathrm{th}}=20\,\mathrm{dBm}$. However, for applied powers in the vicinity of the threshold power $P_{2f,\mathrm{th}}$ the number of parametrically generated magnons during the short pumping pulse is not sufficient to be detected at the measurement point. Consequently, parametrically generated spin waves are only detected if the parametric instability threshold power is overcome by several dB ($P_{2f}>24\,\mathrm{dBm}$). In contrast, if the excitation and the amplification overlap in time, the BLS intensity already starts to increase for powers larger than $P_{2f} = 18\,\mathrm{dBm}$, i.e., below the instability threshold. This is due to the fact that even below this threshold, the signal spin waves exhibit a reduced damping within the amplifier area. In the vicinity of the threshold, this enables a sufficient amount of spin waves to reach the measurement point.  

Comparing the black and the green curve, it is obvious that for rather low $P_{2f}$, the increase in the BLS intensity with $P_{2f}$ is almost independent of the shift between the pulses and it does not really matter whether the spin-wave packet or the pumping pulse arrives first. However, for large $P_{2f} > 24\,\mathrm{dBm}$ the detected intensities in these two cases deviate drastically from each other: If the pumping pulse arrives prior to the traveling spin waves, the BLS intensity does not increase further with $P_{2f}$ and, finally, coincides with the intensity created from parametric generation only, while it continuously increases with $P_{2f}$ if the spin waves arrive first. Since at large powers, there is a large number of parametrically generated spin waves within the amplifier before the spin-wave packet arrives, this is a strong hint that the presence of these waves suppresses the amplification. 

To further analyze the influence of the arrival time of the signal spin-wave packet at the amplifier relative to the beginning of the pumping pulse, the influence of the delay $\Delta t$ is studied more systematically. In Fig.~\ref{Figure2}~(b) the amplification is shown as a function of $\Delta t$ for different powers $P_{2f}$ recorded at the same measurement point as in Fig.~\ref{Figure2}~(a). To estimate the amplification, the measurement noise is extrapolated from Fig.~\ref{Figure1}~(b) and subtracted from the BLS intensity. Subsequently, the amplification is calculated relative to the intensity measured for a delay of $\Delta t = -30\,\mathrm{ns}$, where no temporal overlap between the pulses exists. For this delay, the measured intensity corresponds to the sum of the intensities of two individual, independent spin-wave packets: One excited at the antenna and one generated in the amplifier. Two striking differences between rather large and rather small powers $P_{2f}$ above the threshold are visible: 1) For powers $P_{2f}>23\,\mathrm{dBm}$, the amplification decreases with increasing power $P_{2f}$. Still, the net amount of spin waves reaching the measurement point increases (cf. Fig.~\ref{Figure2}~(a)). 2) While for small powers the amplification is symmetric with respect to $\Delta t = 0\,\mathrm{ns}$, it becomes increasingly asymmetric with increasing $P_{2f}$ and for $P_{2f} = 30\,\mathrm{dBm}$ a notable amplification is only obtained if the spin-wave packet arrives first at the amplifier ($\Delta t > 0$).

In order to understand these dynamics, Fig.~\ref{Figure2}~(c) shows the time-resolved BLS intensity recorded at the same measurement position for an applied power of $P_{2f} = 30\,\mathrm{dBm}$ for different delays $\Delta t$. The extent of the externally excited spin-wave packet in time is indicated by the shaded area. For $\Delta t = -35\,\mathrm{ns}$, the spin-wave packet created by parametric generation occurs at early times $t$ while the spin-wave packet created by the antenna is not visible. This does not change up to a delay of $\Delta t = -5\,\mathrm{ns}$, even though in this case there is quite a substantial overlap between the spin-wave and the pumping pulse. It should be noted that the pumping pulse starts about $5\,\mathrm{ns}$ before the generated spin waves are detected with BLS, since, initially, a sufficient amount of spin waves needs to be generated in order to be detectable. In contrast, if the spin-wave packet arrives prior to the pumping, the profile of the measured pulse changes and a net amplification of the spin-wave packet can be observed. In this case, i.e., for $\Delta t = 0\,\mathrm{ns}$, the BLS intensity rises to an one order of magnitude higher level in comparison to the parametrically generated packet. This intensity level slowly decreases in time due to nonlinear magnon-magnon interactions. It should be noted that the observed effects are in agreement with the observations made in Ref. \onlinecite{Braecher-2014-2}, where especially for large powers a decreased range for an efficient amplification was observed.
 
The strong dependence of the amplification on the arrival time of the spin-wave packet relative to the pumping pulse for large pumping powers can be understood as follows: If the spin-wave packet arrives after the pumping has started, parametric generation rapidly increases the number of thermal spin waves within the amplifier. Thus, at the time the spin-wave packet arrives, there is already a substantial amount of parametrically generated magnons present. The interaction of the externally excited spin waves and the pumping field with these generated magnons suppresses the amplification and limits the output of the amplifier. 

\begin{figure}[t!]
	  \begin{center}
    \scalebox{1}{\includegraphics[width=7.5 cm, clip]{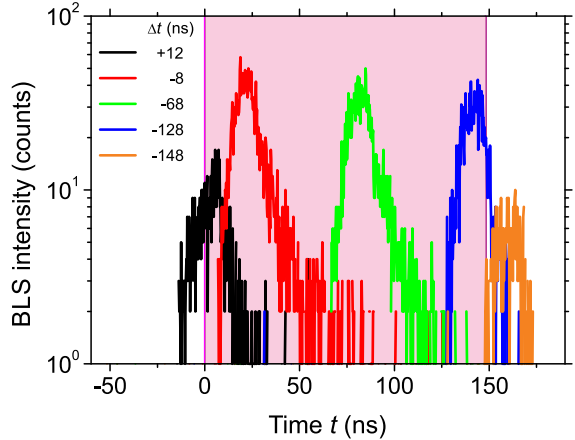}}
    \end{center}
	  \caption{\label{Figure3}Time-resolved BLS intensity for different delays $\Delta t$ for a short spin-wave packet shifted throughout a long pumping pulse with $P_{2f} = 22\,\mathrm{dBm}$ The shaded area marks the position of the pumping pulse in time.}
\end{figure}

Therefore, the parametric amplifier can be operated in two different regimes. In the large power regime, the pumping leads to very high spin-wave intensities and a decent amplification if applied with proper timing (the amplification in the case of $P_{2f} = 30\,\mathrm{dBm}$ in Fig.~\ref{Figure2}~(b) still measures about $10\,\mathrm{dB}$). However, if the spin waves arrive too late, the amplification is suppressed. Thus, in this working regime, the amplifier can be used to filter spin waves in time. On the other hand, if the pumping power is just above the instability threshold, it is possible to operate the amplifier in a quasi CW mode and the spin-wave packet can enter the amplifier at practically any time during the pumping pulse. This case corresponds to the measurements with low powers $P_{2f}$ presented in Fig.~\ref{Figure2}~(b), where the amplification is symmetric with respect to the arrival time of the spin-wave packet in the amplifier. To demonstrate this important working point, Fig.~\ref{Figure3} shows the time-resolved BLS intensity obtained if a spin-wave packet with $\tau_f =15\,\mathrm{ns}$ is shifted in time throughout a $\tau_{2f} = 148\,\mathrm{ns}$ long pumping pulse, whose temporal position is indicated by the shaded area.  The measurement has been performed within the amplifier for an applied pumping power $P_{2f} = 22\,\mathrm{dBm}$. While no parametric generation can be observed, the spin-wave packet is amplified as long as it has a temporal overlap with the pumping pulse. Hereby, the efficiency of the amplification is almost independent of the arrival time, since even for long delays only a very small number of parametrically generated magnons is present in the amplifier when the spin-wave packet arrives.

In conclusion, we have demonstrated the realization of a localized spin-wave amplifier based on the parallel parametric amplification of spin waves in a transversely in-plane magnetized Ni$_{81}$Fe$_{19}$ waveguide. Due to the localization it was possible to study the influence of the parametrically generated magnons on the amplification process. We have shown that the device always works as an efficient amplifier as long as the pumping is timed properly with respect to the arrival time of the spin-wave packet. In the case of a strong pumping, this timing is crucial and the spin waves have to arrive prior to the pumping pulse. This allows for a filtering of spin waves in the time domain. On the other hand, if the applied pumping is rather weak, the timing becomes less important and a higher gain can be achieved. We demonstrated that, in this regime, it is possible to amplify a short spin-wave packet within a long pumping pulse, practically independent of the arrival time of the spin-wave packet within the pumping pulse. This allows for an efficient, frequency selective amplification of spin waves in microstructures for subsequent data processing with a high flexibility for the arrival time.

\begin{acknowledgments}
The authors thank the Nano Structuring Center of the Technische Universit\"at Kaiserslautern for their assistance in sample preparation. T. Br\"acher is supported by a fellowship of the Graduate School Materials Science in Mainz
(MAINZ) through DFG-funding of the Excellence Initiative (GSC 266). Financial support by the DFG (TRR49) is gratefully acknowledged.
\end{acknowledgments}


\begin{thebibliography}{19}

\bibitem{Lenk-2011-1}B.~Lenk, H.~Ulrichs, F.~Garbs, and M.~M\"unzenberg, Physics Reports \textbf{507}, 107 (2011).

\bibitem{Chumak-2010-1}A.~A.~Serga, A.~V.~Chumak, and B.~Hillebrands, J. Phys. D: Appl. Phys. \textbf{43}, 264002 (2010).

\bibitem{Stamps-2014-1}R.~L.~Stamps, S.~Breitkreutz, J.~\r{A}kerman, A.~V. Chumak, Y.~Otani, G.~E.~W. Bauer, J.-U.~Thiele, M.~Bowen, S.~A.~Majetich, M.~Kl\"aui, I.~L.~Prejbeanu, B.~Dieny, N.~M.~Dempsey, and B.~Hillebrands,
J. Phys. D: Appl. Phys. \textbf{47}, 333001 (2014).

\bibitem{Schneider-2008-1}T.~Schneider, A.~A.~Serga, B.~Leven, R.~L.~Stamps, M.~P.~Kostylev, and B. Hillebrands, Appl. Phys. Lett.  \textbf{92}, 022505 (2008).

\bibitem{Chumak-2014-1}A.~A.~Serga, A.~V.~Chumak, and B.~Hillebrands, Nat. Commun. \textbf{5}, 4700 (2014).

\bibitem{Pirro-2011-1}P.~Pirro, T.~Br\"acher, K.~Vogt, B.~Obry, H.~Schultheiss, B.~Leven, and B.~Hillebrands, Phys. Status Solidi B \textbf{248}, 2404 (2011).

\bibitem{Lee-2008-1}K.-S. Lee and S.-K. Kim, J. Appl. Phys. \textbf{104}, 053909 (2008). 

\bibitem{Sebastian-2012-1}T.~Sebastian, Y.~Ohdaira, T.~Kubota, P.~Pirro, T.~Br\"acher, K.~Vogt, A.~A.~Serga, H.~Naganuma, M.~Oogane, Y.~Ando, and B.~Hillebrands, Appl. Phys. Lett. \textbf{100}, 112402 (2012).

\bibitem{Pirro-2014-1}P.~Pirro, T.~Br\"acher, A.~V.~Chumak, B.~L\"agel, C.~Dubs, O.~Surzhenko, P.~G\"ornert, B.~Leven, and B.~Hillebrands, Appl. Phys. Lett. \textbf{104}, 012402 (2014).

\bibitem{Schloemann-1960-1}E.~Schl\"omann, J.~J.~Green, and U.~Milano, J. Appl. Phys. \textbf{31}, 386S (1960).

\bibitem{Braecher-2014-2}T.~Br\"acher, P.~Pirro, T.~Meyer, F.~Heussner, B.~L\"agel, A.~A.~Serga, and B.~Hillebrands, Appl. Phys. Lett. \textbf{104}, 202408 (2014).

\bibitem{Demidov-2009-1}V.~E.~Demidov, M.~P.~Kostylev, K.~Rott, P.~Krzysteczko, G.~Reiss, and S.~O.~Demokritov, Appl. Phys. Lett. \textbf{95}, 112509 (2009).

\bibitem{Melkov-1996-1}A.~G.~Gurevich and G.~A.~Melkov, \textit{Magnetization Oscillations and Waves} (CRC, New York, 1996).

\bibitem{Lvov-1994-1}V.~S.~L'vov, \textit{Wave Turbulence under Parametric Excitation} (Springer-Verlag, Berlin-Heidelberg, 1994).

\bibitem{Smith-2007-1}K.~R.~Smith, V.~I.~Vasyuchka, M.~Wu, G.~A.~Melkov, and C.~E.~Patton, Phys. Rev. B \textbf{76}, 054412 (2007).

\bibitem{Braecher-2011-1}T.~Br\"acher, P.~Pirro, B.~Obry, B.~Leven, A.~A.~Serga, and B.~Hillebrands, Appl. Phys. Lett. \textbf{99}, 162501 (2011).

\bibitem{Demidov-2004-1}V.~E.~Demidov, S.~O.~Demokritov, B.~Hillebrands, M.~Laufenberg, and P.~P.~Freitas, Appl. Phys. Lett. \textbf{85}, 2866 (2004).

\bibitem{Stancil-2009-1}D.~D.~Stancil, A.~Prabhakar, \textit{Spin Waves - Theory and Applications} (Springer, New York, Berlin, Heidelberg, 2009).

\bibitem{Braecher-2014-1}T.~Br\"acher, P.~Pirro, F.~Heussner, A.~A.~Serga, and B.~Hillebrands, Appl. Phys. Lett. \textbf{104}, 092418 (2014).

\end{thebibliography}
\end{document}